\documentclass[conference]{IEEEtran}
\usepackage{multirow}
\usepackage{diagbox}
\usepackage{amssymb}
\usepackage{amsmath}
\usepackage{graphicx}
\usepackage{cite}
\usepackage{citesort}
\usepackage{subfigure}
\usepackage{graphicx,epstopdf}
\usepackage{epsfig}	
\usepackage{cite,graphicx,amsmath,amssymb}
\usepackage{comment}

\usepackage{amssymb}
\usepackage{amsmath}
\usepackage{cite}
\usepackage{url}
\usepackage{xcolor}
\usepackage{cite,graphicx,amsmath,amssymb}
\usepackage{subfigure}
\usepackage{citesort}
\usepackage{fancyhdr}
\usepackage{mdwmath}
\usepackage{mdwtab}
\usepackage{caption}
\usepackage{amsthm}
\usepackage{lipsum}

\newtheorem{theorem}{Theorem}

\newtheorem{lemma}{Lemma}

\newtheorem{corollary}{Corollary}

\newtheorem{remark}{Remark}  
\newtheorem{proposition}{Proposition}

\makeatletter
\def\ScaleIfNeeded{%
\ifdim\Gin@nat@width>\linewidth \linewidth \else \Gin@nat@width
\fi } \makeatother

\begin{document}

\title{\Huge{Outage Performance of Two-Way Relay Non-Orthogonal Multiple Access Systems}}

\author{
\IEEEauthorblockN{  Xinwei~Yue\IEEEauthorrefmark{1}, Yuanwei Liu\IEEEauthorrefmark{2}, Shaoli Kang\IEEEauthorrefmark{1}, Arumugam Nallanathan\IEEEauthorrefmark{2}, and Yue Chen\IEEEauthorrefmark{2}   }

\IEEEauthorblockA{\IEEEauthorrefmark{1}Beihang University, Beijing, China\\
\IEEEauthorrefmark{2} Queen Mary University of London, London, UK\\
 } }

\maketitle

\begin{abstract}
This paper investigates a two-way relay non-orthogonal multiple access (TWR-NOMA) system, where two groups of NOMA users exchange messages with the aid of one half-duplex (HD) decode-and-forward (DF) relay.
Since the signal-plus-interference-to-noise ratios (SINRs) of NOMA signals mainly depend on effective successive interference cancellation (SIC) schemes, imperfect SIC (ipSIC) and perfect SIC (pSIC) are taken into consideration. To characterize the performance of TWR-NOMA systems, we derive closed-form expressions for both exact and asymptotic outage probabilities of NOMA
users' signals with ipSIC/pSIC. Based on the results derived, the diversity order and throughput of the system are examined.
Numerical simulations demonstrate that: 1) TWR-NOMA is superior to TWR-OMA in terms of outage probability in low SNR regimes; and 2) Due to the impact of interference signal (IS) at the relay, error floors and throughput ceilings exist in outage probabilities and ergodic rates for TWR-NOMA, respectively.
\end{abstract}
\section{Introduction}
With the purpose to meet the requirements of future radio access, the design of non-orthogonal multiple access (NOMA) technologies is important to enhance spectral efficiency and user access \cite{Yuanwei2017Proceeding}. The major viewpoint of NOMA is to superpose multiple users by sharing radio resources (i.e., time/frequencey/code) over different power levels \cite{Ding2017Mag,Islam7676258,QinNOMA}. Then the desired signals are detected by exploiting the successive interference cancellation (SIC) \cite{Cover1991Elements}.
Very recently, the integration of cooperative communication with NOMA has been widely discussed in many treaties \cite{Ding2014Cooperative,Liu7445146SWIPT,Men7454773,Yue2016Non}.
Cooperative NOMA has been proposed in \cite{Ding2014Cooperative}, where the user with better channel condition acts as a decode-and-forward (DF) relay to forward information.
With the objective of improving energy efficiency, the application of simultaneous wireless information and power transfer (SWIPT) to the nearby user was investigated where the locations of NOMA users were modeled by stochastic geometry \cite{Liu7445146SWIPT}.
Considering the impact of imperfect channel state information (CSI), the authors in \cite{Men7454773} investigated the performance of amplify-and-forward (AF) relay for downlink NOMA networks, where the exact and tight bounds of outage probability were derived.
To further enhance spectrum efficiency, the performance of full-duplex (FD) cooperative NOMA was characterized in terms of outage behaviors \cite{Yue2016Non}, where user relaying was capable of switching operation between FD and HD mode.

Above existing treaties on cooperative NOMA are all based on one-way relay scheme, where the messages are delivered in only one direction, (i.e., from the BS to the relay or user destinations).
As a further advance, two-way relay (TWR) technique introduced in \cite{Shannon1961Two} has attracted remarkable interest as it is capable of boosting spectral efficiency. The basic idea of TWR systems is to exchange information between two nodes with the help of a relay.
In \cite{Hyadi7004056}, the authors studied the outage behaviors of DF relay with perfect and imperfect CSI conditions.
In terms of CSI and system state information (SSI), the system outage behavior was investigated for two-way full-duplex (FD) DF relay on different multi-user scheduling schemes \cite{Li7778253}.

Motivated by the above two technologies, we focus our attentions on the outage behaviors of TWR-NOMA systems, where two groups of NOMA users exchange messages with the aid of a relay node using DF protocol. Considering
both perfect SIC (pSIC) and imperfect SIC (ipSIC), we derive the closed-form expressions of outage probabilities for users' signals. To provide valuable insights, we further derive the asymptotic outage probabilities of users' signals and obtain the diversity orders.
We show that the outage performance of TWR-NOMA is superior to TWR-OMA in the low signal-to-noise ratio (SNR) regime. We demonstrate that the outage probabilities for TWR-NOMA converge to error floors due to the effect of interference signal (IS) at the relay. We confirm that the use of pSIC is incapable of overcoming the zero diversity order for TWR-NOMA. Additionally, we discuss the system throughput in delay-limited transmission mode.
\section{System Model}\label{System Model}
We consider a two-way relay NOMA communication scenario which consists of one relay $R$, two pairs of NOMA users ${G_1} = \left\{ {{D_1},{D_2}} \right\}$ and ${G_2} =\left\{ {{D_3},{D_4}} \right\}$.
Assuming that $D_{1}$ and $D_{3}$ are the nearby users in group ${G_1}$ and ${G_2}$, respectively, while $D_{2}$ and $D_{4}$ are the distant users in group ${G_1}$ and ${G_2}$, respectively.
The exchange of information between user groups $G_{1}$ and $G_{2}$ is facilitated via the assistance of a decode-and-forward (DF) relay with two antennas, namely $A_1$ and $A_2$. User nodes are equipped with single antenna and can transmit the superposed signals \cite{Ding6868214,Yuanwei2017JSAC}. In addition, we assume that the direct links between two pairs of users are inexistent due to the effect of strong shadowing. Without loss of generality, all the wireless channels are modeled to be independent quasi-static block Rayleigh fading channels and disturbed by additive white Gaussian noise with mean power $N_{0}$. We denote that ${h_{1}}$, ${h_{2}}$, ${h_{3}}$ and ${h_{4}}$ are denoted as the complex channel coefficient of $D_{1}  \leftrightarrow R$, $D_{2} \leftrightarrow R$, $D_3 \leftrightarrow R$ and $D_4 \leftrightarrow R$ links, respectively. The channel power gains ${|h_{1}}|^2$, ${|h_{2}}|^2$, ${|h_{3}}|^2$ and ${|h_{4}}|^2$ are assumed to be exponentially distributed random variables (RVs) with the parameters $ \Omega_{i}$, $\emph{i}\in\{1,2,3,4\}$, respectively. It is assumed that the perfect CSIs of NOMA users are available at $R$ for signal detection.

During the first slot, the pair of NOMA users in $G_1$ transmit the signals to $R$ just as uplink NOMA.
Due to $R$ is equipped with two antennas, when the $R$ receives the signals from the pair of users in $G_1$, it will suffer from
interference signals from the pair of users in $G_2$. More precisely, the observation at $R$ for $A_{1}$ is given by
\begin{align}\label{the expression of R for A1}
{y_{{R_{{A_1}}}}} = {h_1}\sqrt {{a_1}P_u} {x_1} + {h_2}\sqrt {{a_2}P_u} {x_2} + {{\varpi _1}} {I_{{R_{{A_2}}}}} + {n_{{R_{{A_1}}}}},
\end{align}
where ${I_{{R_{{A_2}}}}} $ denotes IS from $A_2$ with ${I_{{R_{{A_2}}}}} = ({h_3}\sqrt {{a_3}P_u} {x_3} + {h_4}\sqrt{{a_4}P_u} {x_4})$. ${\varpi _1} \in \left[ {0,1} \right]$ denotes the impact levels of IS at $R$.
$P_u$ is the normalized transmission power at user nodes.
$x_1$, $x_{2}$ and $x_{3}$, $x_{4}$ are the signals of $D_{1}$, $D_{2}$ and $D_{3}$, $D_{4}$, respectively, i.e,
$\mathbb{E}\{x_{1}^2\}= \mathbb{E}\{x_{2}^2\}=\mathbb{E}\{x_{3}^2\}=\mathbb{E}\{x_{4}^2\}=1$. $a_1$, $a_{2}$ and $a_{3}$, $a_{4}$ are
the corresponding power allocation coefficients. Note that the efficient uplink power control is capable of enhancing the performance of the systems considered, which is beyond the scope of this paper. ${n_{R_{A_{j}}}}$ denotes the Gaussian noise at $R$ for $A_j$, $j \in \left\{ {1,2}
\right\}$.

Similarly, when $R$ receives the signals from the pair of users in $G_2$, it will suffer from interference signals from the pair
of users in $G_1$ as well and then the observation at $R$ is given by
\begin{align}\label{the expression of R for A2}
{y_{{R_{{A_2}}}}} = {h_3}\sqrt {{a_3}P_u} {x_3} + {h_4}\sqrt {{a_4}P_u} {x_4} + {{\varpi _1}} {I_{{R_{{A_1}}}}} + {n_{{R_{{A_2}}}}},
\end{align}
where ${I_{{R_{{A_1}}}}} $ denotes the interference signals from $A_1$ with ${I_{{R_{{A_1}}}}} = ({h_1}\sqrt {{a_1}P_u} {x_1} + {h_2}\sqrt
 {{a_2}P_u} {x_2})$.

Applying the NOMA protocol, $R$ first decodes $D_l$'s information $x_l$ by the virtue of treating $x_t$ as IS. Hence
the received signal-to-interference-plus-noise ratio (SINR) at $R$ to detect $x_l$ is given by
\begin{align}\label{the expression SINR of R to detect x1 or x3}
{\gamma _{R \to {x_l}}} = \frac{{\rho {{\left| {{h_l}} \right|}^2}{a_l}}}{{\rho {{\left| {{h_t}} \right|}^2}{a_t} + \rho {\varpi _1}({{\left| {{h_k}} \right|}^2}{a_k} + {{\left| {{h_r}} \right|}^2}{a_r}) + 1}},
\end{align}
where $\rho  = \frac{{{P_u}}}{{{N_0}}}$ denotes the transmit signal-to-noise ratio (SNR), $\left( {l,k} \right) \in \left\{ {\left( {1,3} \right),\left( {3,1} \right)} \right\}$, $\left( {t,r} \right) \in \left\{ {\left( {2,4} \right),\left( {4,2} \right)} \right\}$.

After SIC is carried out at $R$ for detecting $x_l$, the received SINR at $R$ to detect $x_t$ is given by
\begin{align}\label{the expression SINR of R to detect x2 or x4}
{\gamma _{R \to {x_t}}} = \frac{{\rho {{\left| {{h_t}} \right|}^2}{a_t}}}{{\varepsilon \rho {{\left| g \right|}^2} + \rho {\varpi _1}({{\left| {{h_k}} \right|}^2}{a_k} + {{\left| {{h_r}} \right|}^2}{a_r}) + 1}},
\end{align}
where $\varepsilon = 0$ and $\varepsilon = 1$ denote the pSIC and ipSIC employed at $R$, respectively. Due to the impact of ipSIC, the residual IS is modeled as Rayleigh fading channels \cite{RelaySharing7819537} denoted as $g$ with zero mean and variance ${{\Omega _I}}$.

In the second slot, the information is exchanged between $G_1$ and $G_2$ by the virtue of $R$. Therefore, just like the downlink NOMA, $R$
transmits the superposed signals $( {\sqrt {{b_1}P_r} {x_1} + \sqrt {{b_2}P_r} {x_2}} )$ and $( {\sqrt {{b_3}P_r} {x_3} + \sqrt {{b_4}P_r} {x_4}} )$ to $G_2$ and $G_1$ by $A_2$ and $A_1$, respectively.
$b_1$ and $b_{2}$ denote the power allocation coefficients of $D_1$ and $D_2$, while $b_{3}$ and $b_{4}$ are the corresponding power allocation coefficients of $D_3$ and $D_4$, respectively.
$P_{r}$ is the normalized transmission power at $R$. In particular, to ensure the fairness between users in $G_1$ and $G_2$, a higher power should be allocated to the distant user who has the worse channel conditions. Hence we assume that ${b_2} > {b_1}$ with $b_{1} +b_{2} = 1$ and ${b_4} > {b_3}$ with $b_{3} +b_{4} = 1$. Note that the fixed power allocation coefficients for two groups' NOMA users are considered. Relaxing this assumption will further improve the performance of systems and should be concluded in our future work.

According to NOMA protocol, SIC is employed and the received SINR at $D_k$ to detect $x_t$ is given by
\begin{align}\label{the SINR expression for D1 or D3 to detect x4 or x2}
{\gamma _{{D_k} \to {x_t}}} = \frac{{\rho {{\left| {{h_k}} \right|}^2}{b_t}}}{{\rho {{\left| {{h_k}} \right|}^2}{b_l} + \rho {\varpi _2}{{\left| {{h_k}} \right|}^2} + 1}},
\end{align}
where ${\varpi _2} \in \left[ {0,1} \right]$ denotes the impact level of IS at the user nodes. Then $D_k$ detects $x_l$ and gives the corresponding SINR as follows:
\begin{align}\label{the SINR expression for D1 or D3 to detect its own information}
{\gamma _{{D_k} \to {x_l}}} = \frac{{\rho {{\left| {{h_k}} \right|}^2}{b_l}}}{{\varepsilon \rho {{\left| g \right|}^2} + \rho {\varpi _2}{{\left| {{h_k}} \right|}^2} + 1}}.
\end{align}
Furthermore, the received SINR at $D_t$ to detect $x_r$ is given by
\begin{align}\label{the SINR expression for D2 or D4}
{\gamma _{{D_r} \to {x_t}}} = \frac{{\rho {{\left| {{h_r}} \right|}^2}{b_t}}}{{\rho {{\left| {{h_r}} \right|}^2}{b_l} + \rho {\varpi _2}{{\left| {{h_r}} \right|}^2} + 1}}.
\end{align}

From above process, the exchange of information is achieved between the NOMA users for $G_1$ and $G_2$.
\section{Outage Probability}\label{Section_III}
In this section, the performance of TWR-NOMA is characterized in terms of outage probability.
\subsubsection{Outage Probability of $x_{l}$}\label{the expression of outage for x1}
In TWR-NOMA, the outage events of $x_l$ are explained as follow: i) $R$ cannot decode $x_l$ correctly; ii) The information $x_t$ cannot be detected by $D_k$; and iii) $D_{k}$ cannot detect $x_l$, while $D_{k}$ can first decode $x_t$ successfully.
To simplify the analysis, the complementary events of $x_1$ are employed to express its outage probability. Hence
the outage probability of $x_{l}$ with ipSIC for TWR-NOMA is expressed as
\begin{align}\label{OP expression for x1}
 P_{{x_l}}^{ipSIC} =& 1 - \Pr \left( {{\gamma _{R \to {x_l}}} > {\gamma _{t{h_l}}}} \right) \nonumber \\
  &\times \Pr \left( {{\gamma _{{D_k} \to {x_t}}} > {\gamma _{t{h_t}}},{\gamma _{{D_k} \to {x_l}}} > {\gamma _{t{h_l}}}} \right)  ,
\end{align}
where $\varepsilon = 1$, ${\varpi _1} \in \left[ {0,1} \right]$ and ${\varpi _2} \in \left[ {0,1} \right]$. ${\gamma _{t{h_l}}} = {2^{2{R_l}}} - 1$ with $R_{l}$ being the target rate
at $D_{k}$ to detect $x_{l}$ and ${\gamma _{t{h_t}}} = {2^{2{R_t}}} - 1$ with $R_{t}$ being the target rate at $D_{k}$ to detect $x_{t}$.

The following theorem provides the outage probability of $x_{l}$ for TWR-NOMA.
\begin{theorem} \label{theorem:1 the outage of x1}
The closed-form expression for the outage probability of $x_{l}$ for TWR-NOMA with ipSIC is given by
\begin{align}\label{OP derived for x1}
&P_{{x_l}}^{ipSIC} = 1 - {e^{ - \frac{\beta_l }{{{\Omega _l}}}}}\prod\limits_{i = 1}^3 {{\lambda _i}} \left( {\frac{{{\Phi _1}{\Omega _l}}}{{{\Omega _l}{\lambda _1}{\rm{ + }}\beta_l }} - \frac{{{\Phi _2}{\Omega _l}}}{{{\Omega _l}{\lambda _2}{\rm{ + }}\beta_l }}} \right. \nonumber \\
 &\left. { + \frac{{{\Phi _3}{\Omega _l}}}{{{\Omega _l}{\lambda _3}{\rm{ + }}\beta_l }}} \right)\left( {{e^{ - \frac{\theta_l }{{{\Omega _k}}}}} - \frac{{\varepsilon \tau_l  \rho {\Omega _I}}}{{{\Omega _k} + \varepsilon \rho \tau_l {\Omega _I}}}{e^{ - \frac{{\theta_l \left( {{\Omega _k} + \varepsilon\rho \tau_l  {\Omega _I}} \right)}}{{\varepsilon \tau_l  \rho {\Omega _I}{\Omega _k}}}{\rm{ + }}\frac{1}{{\varepsilon \rho {\Omega _I}}}}}} \right)  ,
\end{align}
where $\varepsilon  = 1$. ${\lambda _1}{\rm{ = }}\frac{1}{{\rho {a_t}{\Omega _t}}}$, ${\lambda _2}{\rm{ = }}\frac{1}{{\rho {\varpi _1}{a_k}{\Omega _k}}}$ and ${\lambda _3}{\rm{ = }}\frac{1}{{\rho {\varpi _1}{a_r}{\Omega _r}}}$. $\beta_l {\rm{ = }}\frac{{{\gamma _{t{h_l}}}}}{{\rho {a_l}}}$. ${\Phi _1}{\rm{ = }}\frac{1}{{\left( {{\lambda _2} - {\lambda _1}} \right)\left( {{\lambda _3} - {\lambda _1}} \right)}}$,${\Phi _2}{\rm{ = }}\frac{1}{{\left( {{\lambda _3} - {\lambda _2}} \right)\left( {{\lambda _2} - {\lambda _1}} \right)}}$ and ${\Phi _3}{\rm{ = }}\frac{1}{{\left( {{\lambda _3} - {\lambda _1}} \right)\left( {{\lambda _3} - {\lambda _2}} \right)}}$. $\theta_l  \buildrel \Delta \over
 = \max \left( {\tau_l ,\xi_t } \right)$.
$\tau_l {\rm{ = }}\frac{{{\gamma _{t{h_l}}}}}{{\rho \left( {{b_l} - {\varpi _2}{\gamma _{t{h_l}}}} \right)}}$ with ${b_l} > {\varpi _2}{\gamma _{t{h_l}}}$ and $\xi_t {\rm{ = }}\frac{{{\gamma _{t{h_t}}}}}{{\rho \left( {{b_t} - {b_l}{\gamma _{t{h_t}}} - {\varpi _2}{\gamma _{t{h_t}}}} \right)}}$ with ${b_t} > \left( {{b_l} + {\varpi _2}} \right){\gamma _{t{h_t}}}$.
\begin{proof}
See Appendix~A.
\end{proof}

\end{theorem}
\begin{corollary}
Based on \eqref{OP derived for x1}, for the special case $\varepsilon = 0$,  the outage probability of $x_{1}$ for TWR-NOMA with pSIC is given by
\begin{align}\label{corollary1 derived for x1 with perfect SIC}
 P_{{x_l}}^{pSIC} =& 1 - {e^{ - \frac{\beta_l }{{{\Omega _l}}} - \frac{\theta_l }{{{\Omega _k}}}}}\prod\limits_{i = 1}^3 {{\lambda _i}} \left( {\frac{{{\Phi _1}{\Omega _l}}}{{{\Omega _l}{\lambda _1}{\rm{ + }}\beta_l }} - \frac{{{\Phi _2}{\Omega _l}}}{{{\Omega _l}{\lambda _2}{\rm{ + }}\beta_l }}} \right. \nonumber \\
 & \left. { + \frac{{{\Phi _3}{\Omega _l}}}{{{\Omega _l}{\lambda _3}{\rm{ + }}\beta_l }}} \right)  .
\end{align}
\end{corollary}
\subsubsection{Outage Probability of $x_{t}$}\label{the expression of outage for x2}
Based on NOMA principle, the complementary events of outage for $x_t$ have the following cases. One of the cases is that $R$ can first
decode the information $x_l$ and then detect $x_t$. Another case is that either of $D_{k}$ and $D_{r}$ can detect $x_t$ successfully.
Hence the outage probability of $x_{t}$ can be expressed as
\begin{align}\label{OP expression for x2}
 P_{{x_t}}^{ipSIC} =& 1 - \Pr \left( {{\gamma _{R \to {x_t}}} > {\gamma _{t{h_t}}},{\gamma _{R \to {x_l}}} > {\gamma _{t{h_l}}}} \right) \nonumber \\
 & \times \Pr \left( {{\gamma _{{D_k} \to {x_t}}} > {\gamma _{t{h_t}}}} \right)\Pr\left( {{\gamma _{{D_r} \to {x_t}}} > {\gamma _{t{h_t}}}} \right) ,
\end{align}
where $\varepsilon = 1$, ${\varpi _1} \in \left[ {0,1} \right]$ and ${\varpi _2} \in \left[ {0,1} \right]$.

The following theorem provides the outage probability of $x_{t}$ for TWR-NOMA.
\begin{theorem} \label{theorem:2 the outage of x2}
The closed-form expression for the outage probability of $x_{t}$ with ipSIC is given by
\begin{align}\label{OP derived for x2}
 &P_{{x_t}}^{ipSIC} = 1 - \frac{{{e^{ - \frac{\beta_l }{{{\Omega _l}}} - {\beta_t}\varphi_t  - \frac{\xi }{{{\Omega _k}}} - \frac{\xi }{{{\Omega _r}}}}}}}{{\varphi_t {\Omega _t}\left( {1 + \varepsilon {\beta_t} \rho \varphi_t {\Omega _I}} \right)\left( {\lambda _2^{'} - \lambda _1^{'}} \right)}}\prod\limits_{i = 1}^2 {\lambda _i^{'}}  \nonumber\\
 &  \times \left( {\frac{{{\Omega _l}}}{{\beta_l  + {\beta_t}{\Omega _1}\varphi_t  + {\Omega _l}\lambda _1^{'}}} - \frac{{{\Omega _l}}}{{\beta_l  + {\beta_t}{\Omega _1}\varphi_t  + {\Omega _l}\lambda _2^{'}}}} \right),
\end{align}
where $\varepsilon = 1$. $\lambda _1^{'}{\rm{ = }}\frac{1}{{\rho {\varpi _1}{a_k}{\Omega _k}}}$ and $\lambda _2^{'}{\rm{ = }}\frac{1}{{\rho {\varpi _1}{a_r}{\Omega _r}}}$. ${\beta_t} = \frac{{{\gamma _{t{h_t}}}}}{{\rho {a_t}}}$, $\varphi_t  = \frac{{{\Omega _l}+ \rho \beta_l {a_t}{\Omega _t}}}{{{\Omega _l}{\Omega _t}}}$.
\begin{proof}
See Appendix~B.
\end{proof}
\end{theorem}

\begin{corollary}
For the special case, substituting $\varepsilon  = 0$ into \eqref{OP derived for x2}, the outage probability of $x_{2}$ for TWR-NOMA with pSIC is
given by
\begin{align}\label{corollary2 derived for x2 with perfect SIC}
 &P_{{x_t}}^{pSIC} = 1 - \frac{{{e^{ - \frac{\beta_l }{{{\Omega _l}}} - {\beta_t}\varphi_t  - \frac{\xi }{{{\Omega _k}}} - \frac{\xi }{{{\Omega _r}}}}}}}{{\varphi_t {\Omega _t}\left( {\lambda _2^{'} - \lambda _1^{'}} \right)}}\prod\limits_{i = 1}^2 {\lambda _i^{'}} \nonumber \\
  & \times \left( {\frac{{{\Omega _l}}}{{\beta_l  + {\beta_t}{\Omega _l}\varphi_t  + {\Omega _l}\lambda _1^{'}}} - \frac{{{\Omega _l}}}{{\beta_l  + {\beta_t}{\Omega _l}\varphi_t  + {\Omega _l}\lambda _2^{'}}}} \right) .
\end{align}
\end{corollary}
\subsubsection{Diversity Order Analysis}
To obtain deeper insights for TWR-NOMA systems, the asymptotic analysis are presented in high SNR regimes based on the derived outage probabilities. The diversity order is defined as \cite{Liu2016TVT,Yuanwei2017TWC}
\begin{align}\label{diversity order}
d =  - \mathop {\lim }\limits_{\rho  \to \infty } \frac{{\log \left( {P_{x_i}^\infty \left( \rho  \right)} \right)}}{{\log \rho }},
\end{align}
where ${P_{x_i}^\infty }$ denotes the asymptotic outage probability of $x_{i}$.

\begin{proposition}\label{proposition:diversity total for x_l}
Based on the analytical results in \eqref{OP derived for x1} and \eqref{corollary1 derived for x1 with perfect SIC}, when $\rho  \to \infty $, the asymptotic outage probabilities of $x_l$ for ipSIC/pSIC with ${e^{ - x}}\approx  1 - x$ are given by
\begin{align}\label{the asymptotic OP of x1 with ipSIC}
 &P_{{x_l},\infty }^{ipSIC} = 1 - \prod\limits_{i = 1}^3 {{\lambda _i}} \left( {\frac{{{\Phi _1}{\Omega _l}}}{{{\Omega _l}{\lambda _1}{\rm{ + }}{\beta _l}}} - \frac{{{\Phi _2}{\Omega _l}}}{{{\Omega _l}{\lambda _2}{\rm{ + }}{\beta _l}}} + \frac{{{\Phi _3}{\Omega _l}}}{{{\Omega _l}{\lambda _3}{\rm{ + }}{\beta _l}}}} \right) \nonumber \\
  &\times \left[ {1 - \frac{{{\theta _l}}}{{{\Omega _k}}} - \frac{{\varepsilon \tau \rho {\Omega _I}}}{{{\Omega _k} + \varepsilon \rho \tau {\Omega _I}}}\left( {1 - \frac{{{\theta _l}\left( {{\Omega _k} + \varepsilon \tau \rho {\Omega _I}} \right)}}{{\tau \varepsilon \rho {\Omega _I}{\Omega _k}}}} \right)} \right] ,
\end{align}
and
\begin{align}\label{the asymptotic OP of x1 with pSIC}
P_{{x_l},\infty }^{pSIC} = 1 - \prod\limits_{i = 1}^3 {{\lambda _i}} \left( {\frac{{{\Phi _1}{\Omega _l}}}{{{\Omega _l}{\lambda _1}{\rm{ + }}{\beta _l}}} - \frac{{{\Phi _2}{\Omega _l}}}{{{\Omega _l}{\lambda _2}{\rm{ + }}{\beta _l}}} + \frac{{{\Phi _3}{\Omega _l}}}{{{\Omega _l}{\lambda _3}{\rm{ + }}{\beta _l}}}} \right),
\end{align}
respectively. Substituting \eqref{the asymptotic OP of x1 with ipSIC} and \eqref{the asymptotic OP of x1 with pSIC} into \eqref{diversity order}, the diversity orders of $x_l$ with ipSIC/pSIC are equal to zeros.
\end{proposition}

\begin{remark}\label{remark:1}
An important conclusion from above analysis is that due to impact of residual interference, the diversity order of $x_{l}$ with the use of ipSIC is zero. Additionally, the communication process of the first slot similar to uplink NOMA, even though under the condition of pSIC, diversity order is equal to zero as well for $x_{l}$. As can be observed that there are error floors for $x_{l}$ with ipSIC/pSIC.
\end{remark}
\begin{proposition}\label{proposition:diversity total for x_l}
Similar to the resolving process of \eqref{the asymptotic OP of x1 with ipSIC} and \eqref{the asymptotic OP of x1 with pSIC}, the asymptotic outage probabilities of $x_t$ with ipSIC/pSIC in high SNR regimes are given by
\begin{align}\label{the asymptotic OP of x2 with ipSIC}
 &P_{{x_t},\infty }^{ipSIC} = 1 - \frac{{\lambda _1^{'}\lambda _2^{'}}}{{{\varphi _t}{\Omega _t}\left( {1 + \varepsilon \rho {\beta _t}{\varphi _t}{\Omega _I}} \right)\left( {\lambda _2^{'} - \lambda _1^{'}} \right)}} \nonumber \\
  &\times \left( {\frac{{{\Omega _l}}}{{{\beta _l} + {\beta _t}{\Omega _1}{\varphi _t} + {\Omega _l}\lambda _1^{'}}} - \frac{{{\Omega _l}}}{{{\beta _l} + {\beta _t}{\Omega _1}{\varphi _t} + {\Omega _l}\lambda _2^{'}}}} \right)  ,
 \end{align}
and
\begin{align}\label{the asymptotic OP of x2 with pSIC}
 &P_{{x_t},\infty }^{pSIC} = 1 - \frac{{\lambda _1^{'}\lambda _2^{'}}}{{{\varphi _t}{\Omega _t}\left( {\lambda _2^{'} - \lambda _1^{'}} \right)}}\nonumber \\
  & \times \left( {\frac{{{\Omega _l}}}{{{\beta _l} + {\beta _t}{\Omega _1}{\varphi _t} + {\Omega _l}\lambda _1^{'}}} - \frac{{{\Omega _l}}}{{{\beta _l} + {\beta _t}{\Omega _l}{\varphi _t} + {\Omega _l}\lambda _2^{'}}}} \right) ,
\end{align}
respectively. Substituting \eqref{the asymptotic OP of x2 with ipSIC} and \eqref{the asymptotic OP of x2 with pSIC} into \eqref{diversity order}, the diversity orders of $x_{t}$ for both ipSIC and pSIC are zeros.
\end{proposition}
\begin{remark}\label{remark:2}
Based on above analytical results of $x_{l}$, the diversity orders of $x_{t}$ with ipSIC/pSIC are also equal to zeros.
This is because residual interference is existent in the total communication process.
\end{remark}
\subsubsection{Throughput Analysis}\label{Seciton-A-3}
In delay-limited transmission scenario, the BS transmits message to users at a fixed rate, where system throughput will be subject to wireless fading channels. Hence the corresponding throughput of TWR-NOMA with ipSIC/pSIC is calculated as \cite{Liu7445146SWIPT}
\begin{align}\label{delay-limited throughput}
 R_{dl}^{\psi} =& \left( {1 - P_{{x_1}}^{\psi} } \right){R_{{x_1}}} + \left( {1 - P_{{x_2}}^{\psi} } \right){R_{{x_2}}} \nonumber \\
  &+ \left( {1 - P_{{x_3}}^{\psi} } \right){R_{{x_3}}} + \left( {1 - P_{{x_4}}^{\psi} } \right){R_{{x_4}}},
\end{align}
where $\psi  \in \left( {ipSIC,pSIC} \right)$. $P_{{x_1}}^{\psi}$ and $P_{{x_3}}^{\psi}$ with ipSIC/pSIC can be obtained from \eqref{OP derived for x1} and  \eqref{corollary1 derived for x1 with perfect SIC}, respectively, while
$P_{{x_2}}^{\psi}$ and $P_{{x_4}}^{\psi}$ with ipSIC/pSIC can be obtained from  \eqref{OP derived for x2} and \eqref{corollary2 derived for x2 with perfect SIC}, respectively.

\begin{table}[!t]
\centering
\caption{Table of Parameters for Numerical Results}
\tabcolsep5pt
\renewcommand\arraystretch{1} 
\begin{tabular}{|l|l|}
\hline
Monte Carlo simulations repeated  &  ${10^6}$ iterations \\
\hline
\multirow{2}{*}{Power allocation coefficients of NOMA} &  \multirow{1}{*}{ $b_1=b_3=0.2$}   \\
                                                       &  \multirow{1}{*}{ $b_2=b_4=0.8$}   \\
\hline
\multirow{2}{*}{Targeted data rates}  & \multirow{1}{*}{$R_{{1}}=R_{{3}}=0.1 $ BPCU}  \\
                                      & \multirow{1}{*}{$R_{{2}}=R_{{4}}=0.01$ BPCU}  \\
\cline{1-2}
Pass loss exponent  & $\alpha=2$  \\
\hline
The distance between R and $D_{1}$ or $D_{3}$ &  $d_1=2$ m \\
\hline
The distance between R and $D_{2}$ or $D_{4}$ & $d_2=10$ m \\
\hline
\end{tabular}
\label{parameter}
\end{table}
\section{Numerical Results}\label{Numerical Results}
In this section, numerical results are provide to investigate the impact levels of IS on outage probability for TWR-NOMA systems. The simulation parameters used are summarized in Table~\ref{parameter}, where BPCU is short for bit per channel use. Due to the reciprocity of channels between $G_1$ and $G_2$, the outage behaviors of $x_1$ and $x_2$ in $G_1$ are presented to illustrate availability of TWR-NOMA. Without loss of generality, the power allocation coefficients of $x_1$ and $x_2$ are set as $a_1=0.8$ and $a_2=0.2$, respectively. ${\Omega _1}$ and ${\Omega _2}$ are set to be ${\Omega _1} = d_1^{ - \alpha }$ and ${\Omega _2} = d_2^{ - \alpha }$, respectively.

\subsection{Outage Probability}
\begin{figure}[t!]
    \begin{center}
        \includegraphics[width=3.3in,  height=2.4in]{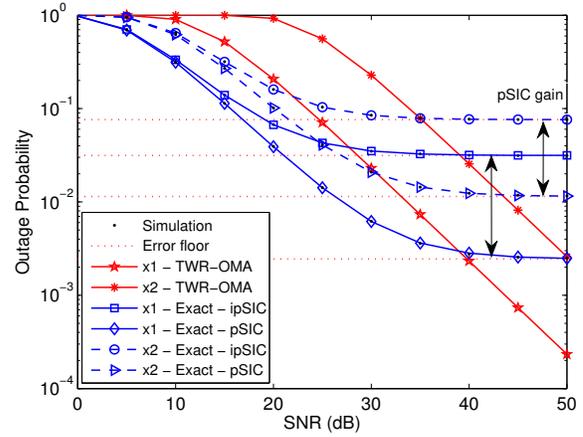}
        \caption{Outage probability versus the transmit SNR.}
        \label{Pout_x1_x2_ipSIC_pSIC_vs_OMA}
    \end{center}
\end{figure}

Fig. \ref{Pout_x1_x2_ipSIC_pSIC_vs_OMA} plots the outage probabilities of $x_1$ and $x_2$ with both ipSIC and pSIC versus SNR for simulation setting with ${\varpi _1} = {\varpi _2}=0.01$ and ${\Omega _I}=-20$ dB. The solid and dashed curves represent the exact theoretical performance of $x_1$ and $x_2$ for both ipSIC and pSIC, corresponding to the results derived in \eqref{OP derived for x1}, \eqref{corollary1 derived for x1 with perfect SIC} and \eqref{OP derived for x2}, \eqref{corollary2 derived for x2 with perfect SIC}, respectively. Apparently, the outage probability curves match perfectly with Monte Carlo simulation results. As can be observed from the figure, the outage behaviors of $x_1$ and $x_2$ for TWR-NOMA are superior to TWR-OMA in the low SNR regime.
This is due to the fact that the influence of IS is not the dominant factor at low SNR.
Furthermore, another observation is that the pSIC is capable of enhancing the performance of NOMA compare to the ipSIC. In addition, the asymptotic curves of $x_1$ and $x_2$ with ipSIC/pSIC are plotted according to \eqref{the asymptotic OP of x1 with ipSIC}, \eqref{the asymptotic OP of x1 with pSIC} and \eqref{the asymptotic OP of x2 with ipSIC}, \eqref{the asymptotic OP of x1 with pSIC}, respectively. It can be seen that the outage behaviors of $x_1$ and $x_2$ converge to the error floors in the high SNR regime. The reason can be explained that due to the impact of residual interference by the use of ipSIC, $x_1$ and $x_2$ result in zero diversity orders. Although the pSIC is carried out in TWR-NOMA system, $x_1$ and $x_2$ also obtain zero diversity orders. This is due to the fact that when the relay first detect the strongest signal in the first slot, it will suffer interference from the weaker signal. This observation verifies the conclusion \textbf{Remark \ref{remark:1}} in Section \ref{Section_III}.
\begin{figure}[t!]
    \begin{center}
        \includegraphics[width=3.3in,  height=2.4in]{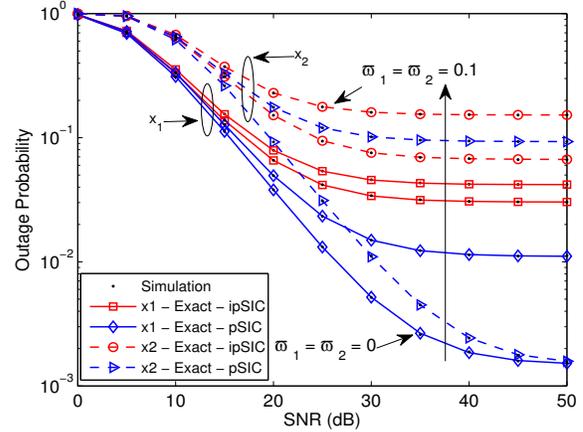}
        \caption{Outage probability versus the transmit SNR, with ${\Omega _I}=-20$ dB.}
        \label{Pout_x1_x2_ipSIC_pSIC_diff_LI}
    \end{center}
\end{figure}

\begin{figure}[t!]
    \begin{center}
        \includegraphics[width=3.3in,  height=2.4in]{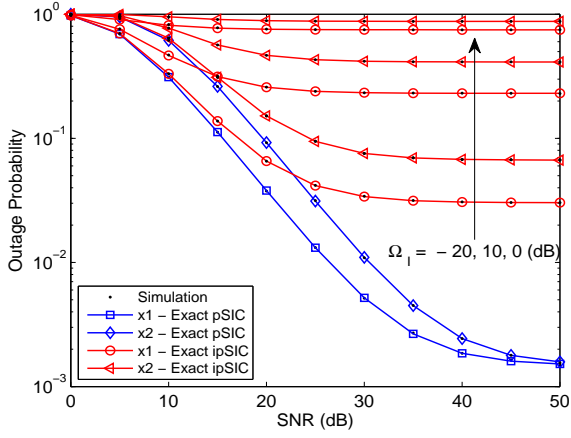}
        \caption{Outage probability versus the transmit SNR, with ${\varpi _1} = {\varpi _2} = 0$.}
        \label{Outage_x1_x2_with_diff_LI_ipSIC}
    \end{center}
\end{figure}

Fig. \ref{Pout_x1_x2_ipSIC_pSIC_diff_LI} plots the outage probabilities of $x_1$ and $x_2$ versus SNR with the different impact levels of IS from ${\varpi _1} = {\varpi _2} = 0$ to ${\varpi _1} = {\varpi _2} = 0.1$.  The solid and dashed curves represent
the outage behaviors of $x_1$ and $x_2$ with ipSIC/pSIC, respectively.
As can be seen that when the impact level of IS is set to be ${\varpi _1} = {\varpi _2} = 0$, there is no IS between $A_1$ and $A_2$ at the relay, which can be viewed as a benchmark. Additionally, one can observed that with the impact levels of IS increasing, the outage performance of TWR-NOMA system degrades significantly. Hence it is crucial to hunt for efficient strategies for suppressing the effect of interference between antennas. Fig. \ref{Outage_x1_x2_with_diff_LI_ipSIC} plots the outage probability versus SNR with different values of residual IS from $-20$ dB to $0$ dB. It can be seen that the different values of residual IS affects the
performance of ipSIC seriously. Similarly, as the values of residual IS increases, the preponderance of ipSIC is inexistent. When ${\Omega _I}=0$ dB, the outage probability of $x_1$ and $x_2$ will be in close proximity to one. Therefore, it is important to design effective SIC schemes for TWR-NOMA.
\begin{figure}[t!]
    \begin{center}
        \includegraphics[width=3.3in,  height=2.4in]{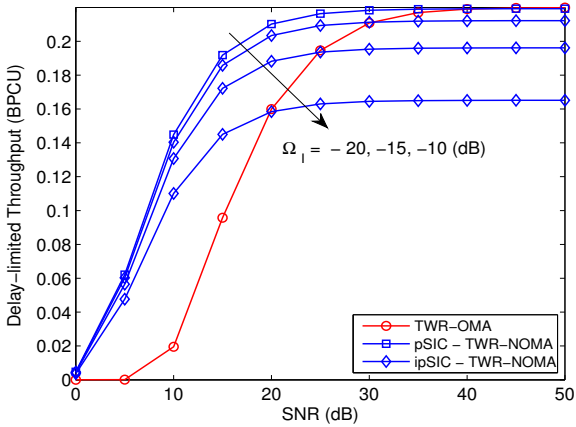}
        \caption{System throughput in delay-limited transmission mode versus SNR with ipSIC/pSIC,
        ${\varpi _1} = {\varpi _2} = 0.01$.}
        \label{delay_limited_throughput}
    \end{center}
\end{figure}

Fig. \ref{delay_limited_throughput} plots system throughput versus SNR in delay-limited transmission mode for TWR-NOMA with different values of residual IS from $-20$ dB to $-10$ dB.
The blue solid curves represent throughput for TWR-NOMA with both pSIC and ipSIC, which can be obtained from \eqref{delay-limited throughput}. One can observe that TWR-NOMA is capable of achieving a higher throughput compared to TWR-OMA in the low SNR regime, since it has a lower outage probability. Moreover, the figure confirms that TWR-NOMA converges to the throughput ceiling in high SNR regimes. It is worth noting that ipSIC considered for TWR-NOMA will further degrade throughput with the values of residual IS becomes larger in high SNR regimes.
\section{Conclusion}\label{Conclusion}
This paper has investigated the application of TWR to NOMA systems, in which two pairs of users can exchange their information between each other by the virtue of a relay node. The performance of TWR-NOMA has been characterized in terms of outage probability and ergodic rate for both ipSIC and pSIC. Furthermore, the closed-form expressions of outage probability for the NOMA users' signals have been derived. Owing to the impact of IS at relay, there were the error floors for TWR-NOMA with ipSIC/pSIC in high SNR regimes and zero diversity orders were obtained. Based on the analytical results, it was shown that the performance of TWR-NOMA with ipSIC/pSIC outperforms TWR-OMA in the low SNR regime.
\appendices
\section*{Appendix~A: Proof of Theorem \ref{theorem:1 the outage of x1}} \label{Appendix:A}
\renewcommand{\theequation}{A.\arabic{equation}}
\setcounter{equation}{0}

Substituting \eqref{the expression SINR of R to detect x1 or x3}, \eqref{the SINR expression for D1 or D3 to detect x4 or x2} and
\eqref{the SINR expression for D1 or D3 to detect its own information} into \eqref{OP expression for x1}, the outage probability of $x_l$
can be further given by
\begin{align}\label{the proof of outage probability for x1 with imperfect SIC}
 &P_{{x_l}}^{ipSIC} = 1 \nonumber \\
  &- \underbrace {\Pr \left( {\frac{{\rho {{\left| {{h_l}} \right|}^2}{a_l}}}{{\rho {{\left| {{h_t}} \right|}^2}{a_t} + \rho {\varpi _1}({{\left| {{h_k}} \right|}^2}{a_k} + {{\left| {{h_r}} \right|}^2}{a_r}) + 1}} > {\gamma _{t{h_l}}}} \right)}_{{J_1}}\nonumber \\
 & \times \Pr \left( {\frac{{\rho {{\left| {{h_k}} \right|}^2}{b_t}}}{{\rho {{\left| {{h_k}} \right|}^2}{b_l} + \rho {\varpi _2}{{\left| {{h_k}} \right|}^2} + 1}} > {\gamma _{t{h_t}}},} \right. \nonumber\\
&\underbrace {\begin{array}{*{20}{c}}
   {} & {} & {} & {} & {} & {}  \\
\end{array}\left. {\frac{{\rho {{\left| {{h_k}} \right|}^2}{b_l}}}{{\varepsilon \rho {{\left| g \right|}^2} + \rho {\varpi _2}{{\left| {{h_k}} \right|}^2} + 1}} > {\gamma _{t{h_l}}}} \right)}_{{J_2}}  ,
\end{align}
where $\varepsilon =1$.

To calculate the probability $J_1$ in \eqref{the proof of outage probability for x1 with imperfect SIC},
let $Z = \rho {a_t}{\left| {{h_t}} \right|^2} + \rho {\varpi _1}{a_k}{\left| {{h_k}} \right|^2} + \rho {\varpi _1}{a_r}{\left| {{h_r}} \right|^2}$. We first calculate the PDF of $Z$ and then give the process derived of $J_{1}$. As is known, ${\left| {{h_i}}
\right|^2}$ follows the exponential distribution with the means ${\Omega _i}$, $i \in \left( {1,2,3,4} \right)$. Furthermore, we denote
that ${Z_1} = \rho {a_t}{\left| {{h_t}} \right|^2}$, ${Z_2} = \rho {\varpi _1}{a_k}{\left| {{h_k}} \right|^2}$ and ${Z_3} = \rho {\varpi _1}{a_r}{\left| {{h_r}} \right|^2}$ are also independent exponentially distributed random variables (RVs) with means ${\lambda _1}{\rm{ = }}\frac{1}{{\rho {a_t}{\Omega _t}}}$, ${\lambda _2}{\rm{ = }}\frac{1}{{\rho {\varpi _1}{a_k}{\Omega _k}}}$ and ${\lambda _3}{\rm{ = }}\frac{1}{{\rho {\varpi _1}{a_r}{\Omega _r}}}$, respectively. Based on \cite{Nadarajah6831670}, for the independent non-identical distributed (i.n.d) fading
scenario, the PDF of $Z$ can be given by
\begin{align}\label{PDF of Z}
{f_Z}\left( z \right){\rm{ = }}\prod\limits_{i = 1}^3 {{\lambda _i}} \left( {{\Phi _1}{e^{ - {\lambda _1}z}} -
{\Phi _2}{e^{ - {\lambda _2}z}}{\rm{ + }}{\Phi _3}{e^{ - {\lambda _3}z}}} \right),
\end{align}
where ${\Phi _1}{\rm{ = }}\frac{1}{{\left( {{\lambda _2} - {\lambda _1}} \right)\left( {{\lambda _3} - {\lambda _1}}
\right)}}$, ${\Phi _2}{\rm{ = }}\frac{1}{{\left( {{\lambda _3} - {\lambda _2}} \right)\left( {{\lambda _2} - {\lambda _1}} \right)}}$ and ${\Phi _3}{\rm{ = }}\frac{1}{{\left( {{\lambda _3} - {\lambda _1}} \right)\left( {{\lambda _3} - {\lambda _2}} \right)}}$.

According to the above explanations, $J_1$ is calculated as follows:
\begin{align}\label{derived process for J1}
{J_1} = \Pr \left( {{{\left| {{h_l}} \right|}^2} > \left( {Z + 1} \right)\beta_l } \right) = \int_0^\infty  {{f_Z}\left( z \right)} {e^{ - \frac{{\left( {z + 1} \right)\beta_l }}{{{\Omega _l}}}}}dz.
\end{align}
Substituting \eqref{PDF of Z} into \eqref{derived process for J1} and after some algebraic manipulations, $J_1$ is given by
\begin{align}\label{the derived expression of J1 finally}
{J_1} = {e^{ - \frac{\beta_l }{{{\Omega _l}}}}}\prod\limits_{i = 1}^3 {{\lambda _i}} \left( {\frac{{{\Phi _1}{\Omega _l}}}{{{\Omega _l}{\lambda _1}{\rm{ + }}\beta_l }} - \frac{{{\Phi _2}{\Omega _l}}}{{{\Omega _l}{\lambda _2}{\rm{ + }}\beta_l }} + \frac{{{\Phi _3}{\Omega _l}}}{{{\Omega _l}{\lambda _3}{\rm{ + }}\beta_l }}} \right),
\end{align}
where $\beta_l {\rm{ = }}\frac{{{\gamma _{t{h_l}}}}}{{\rho {a_l}}}$.

$J_2$ can be further calculated as follows:
\begin{align*}\label{the derived process of J2}
 {J_2} 
  =& \Pr \left( {{{\left| {{h_k}} \right|}^2} > \max \left( {\tau_l ,\xi_t } \right) \buildrel \Delta \over = \theta_l ,{{\left| g \right|}^2} < \frac{{{{\left| {{h_k}} \right|}^2} - \tau_l }}{{\varepsilon \rho \tau_l }}} \right) \nonumber\\
    \end{align*}
  \begin{align}
  =& \int_\theta ^\infty  {\frac{1}{{{\Omega _k}}}} \left( {{e^{ - \frac{y}{{{\Omega _k}}}}} - {e^{ - \frac{{y - \tau_l }}{{\varepsilon \tau_l \rho {\Omega _I}}} - \frac{y}{{{\Omega _k}}}}}} \right)dy \nonumber\\
  =& {e^{ - \frac{\theta_l }{{{\Omega _k}}}}} - \frac{{\tau_l \varepsilon \rho {\Omega _I}}}{{{\Omega _k} + \varepsilon \rho \tau_l {\Omega _I}}}{e^{ - \frac{{\theta_l \left( {{\Omega _k} + \rho \tau_l \varepsilon {\Omega _I}} \right)}}{{\tau_l \varepsilon \rho {\Omega _I}{\Omega _k}}}{\rm{ + }}\frac{1}{{\varepsilon \rho {\Omega _I}}}}} ,
\end{align}
where $\xi_t {\rm{ = }}\frac{{{\gamma _{t{h_t}}}}}{{\rho \left( {{b_t} - {b_l}{\gamma _{t{h_t}}} - {\varpi _2}{\gamma _{t{h_t}}}} \right)}}$ with ${b_t} > \left( {{b_l} + {\varpi _2}} \right){\gamma _{t{h_t}}}$, $\tau_l {\rm{ = }}\frac{{{\gamma _{t{h_l}}}}}{{\rho \left( {{b_l} - {\varpi _2}{\gamma _{t{h_l}}}} \right)}}$ with ${b_l} > {\varpi _2}{\gamma _{t{h_l}}}$.
Combining \eqref{the derived expression of J1 finally} and \eqref{the derived process of J2}, we can obtain \eqref{OP derived for x1}.
The proof is complete.

\appendices
\section*{Appendix~B: Proof of Theorem \ref{theorem:2 the outage of x2}} \label{Appendix:B}
\renewcommand{\theequation}{B.\arabic{equation}}
\setcounter{equation}{0}

Substituting \eqref{the expression SINR of R to detect x1 or x3}, \eqref{the expression SINR of R to detect x2 or x4}, \eqref{the SINR expression for D1 or D3 to detect its own information} and \eqref{the SINR expression for D2 or D4} into \eqref{OP expression for x2}, the outage probability of $x_t$ is rewritten as
\begin{align}\label{the proof of outage probability for x2 with perfect SIC}
&P_{{x_t}}^{ipSIC} = 1 \nonumber \\
&- \Pr \left( {\frac{{\rho {{\left| {{h_t}} \right|}^2}{a_t}}}{{\varepsilon \rho {{\left| g \right|}^2} + \rho {\varpi _1}({{\left| {{h_k}} \right|}^2}{a_k} + {{\left| {{h_r}} \right|}^2}{a_r}) + 1}} > {\gamma _{t{h_t}}},} \right.\nonumber \\
& \begin{array}{*{20}{c}}
   {}  \\
\end{array}\underbrace {\left. {\frac{{\rho {{\left| {{h_l}} \right|}^2}{a_l}}}{{\rho {{\left| {{h_t}} \right|}^2}{a_t} + \rho {\varpi _1}({{\left| {{h_k}} \right|}^2}{a_k} + {{\left| {{h_r}} \right|}^2}{a_r}) + 1}} > {\gamma _{t{h_l}}}} \right)}_{{\Theta _1}} \nonumber\\
 & \times \underbrace {\Pr \left( {\frac{{\rho {{\left| {{h_k}} \right|}^2}{b_t}}}{{\rho {{\left| {{h_k}} \right|}^2}{b_l} + \rho {\varpi _2}{{\left| {{h_k}} \right|}^2} + 1}} > {\gamma _{t{h_t}}}} \right)}_{{\Theta _2}} \nonumber\\
 & \times \underbrace {\Pr \left( {\frac{{\rho {{\left| {{h_r}} \right|}^2}{b_t}}}{{\rho {{\left| {{h_r}} \right|}^2}{b_l} + \rho {\varpi _2}{{\left| {{h_r}} \right|}^2} + 1}} > {\gamma _{t{h_t}}}} \right)}_{{\Theta _3}} ,
\end{align}
where ${\varpi _1} = {\varpi _2} \in \left[ {0,1} \right]$  and $\varepsilon  = 1$.

Similar to \eqref{PDF of Z}, let ${Z^{'}}{\rm{ = }}\rho {\varpi _1}{a_k}{\left| {{h_k}} \right|^2} + \rho {\varpi _1}{a_r}{\left| {{h_r}} \right|^2}$,  the PDF of ${{Z^{'}}}$ is given by
\begin{align}\label{PDF of Z with two varibles}
{f_{{Z^{'}}}}\left( {{z}}^{'} \right) = \prod\limits_{i = 1}^2 {\lambda _i^{'}} \left( {\frac{{{e^{ - \lambda _1^{'}{{z^{'}}}}}}}{{\left( {\lambda _2^{'} - \lambda _1^{'}}
 \right)}} - \frac{{{e^{ - \lambda _2^{'}{{z^{'}}}}}}}{{\left( {\lambda _2^{'} - \lambda _1^{'}} \right)}}} \right),
\end{align}
where $\lambda _1^{'}{\rm{ = }}\frac{1}{{\rho {\varpi _1}{a_k}{\Omega _k}}}$ and $\lambda _2^{'}{\rm{ = }}\frac{1}
{{\rho {\varpi _1}{a_r}{\Omega _r}}}$.

After some variable substitutions and manipulations,
\begin{align}\label{the derived process of Theta1}
 {\Theta _1} =& \Pr \left( {{{\left| {{h_t}} \right|}^2} > {\beta_t}\left( {\varepsilon \rho {{\left| g \right|}^2} + {Z^{'}} + 1} \right),} \right. \nonumber \\
 &\begin{array}{*{20}{c}}
   {} & {} & {} & {} & {\left. {{{\left| {{h_l}} \right|}^2} > \beta_l \left( {\rho {{\left| {{h_t}} \right|}^2}{a_t} + {Z^{'}} + 1} \right)} \right)}  \nonumber \\
\end{array} \nonumber \\
  =& \frac{1}{{\varphi_t {\Omega _t}\left( {1 + \varepsilon \rho{\beta_t}    \varphi_t {\Omega _I}} \right)}}{e^{ - \frac{\beta_l }{{{\Omega _1}}} - {\beta_t}\varphi_t }} \nonumber\\
  &\times \int_0^\infty  {{f_{Z^{'}}}\left( z^{'} \right){e^{ - \frac{{\left( {\beta_l  + {\beta_t}{\Omega _l}\varphi_t } \right)z^{'}}}{{{\Omega _l}}}}}} dx ,
\end{align}
where ${\beta_t} = \frac{{{\gamma _{t{h_t}}}}}{{\rho {a_t}}}$ and $\varphi_t  = \frac{{{\Omega _l} + \rho \beta_l {a_t}{\Omega _t}}}
{{{\Omega _l}{\Omega _t}}}$.

Substituting \eqref{PDF of Z with two varibles} into \eqref{the derived process of Theta1}, ${\Theta _1}$ can be given by
\begin{align}\label{the expression of Theta1}
 &{\Theta _1} = \frac{{{e^{ - \frac{\beta_l }{{{\Omega _l}}} - {\beta_t}\varphi_t }}}}{{\varphi_t {\Omega _t}\left( {1 + {\beta_t}\varepsilon \rho \varphi_t {\Omega _I}} \right)\left( {\lambda _2^{'} - \lambda _1^{'}} \right)}} \nonumber\\
  &\times \prod\limits_{i = 1}^2 {\lambda _i^{'}} \left( {\frac{{{\Omega _l}}}{{\beta_l  + {\beta_t}{\Omega _l}\varphi_t  + {\Omega _l}\lambda _1^{'}}} - \frac{{{\Omega _l}}}{{\beta_l  + {\beta_t}{\Omega _l}\varphi_t  + {\Omega _l}\lambda _2^{'}}}} \right) .
\end{align}

${\Theta _2}$ and ${\Theta _3}$ can be easily calculated
\begin{align}\label{the expression of Theta2}
 {\Theta _2} = \Pr \left( {{{\left| {{h_k}} \right|}^2} > \xi_t } \right){\rm{ = }}{e^{ - \frac{\xi_t }{{{\Omega _k}}}}},
\end{align}
and
\begin{align}\label{the expression of Theta3}
{\Theta _3} = \Pr \left( {{{\left| {{h_r}} \right|}^2} > \xi_t } \right) = {e^{ - \frac{\xi_t }{{{\Omega _r}}}}},
\end{align}
respectively, where $\xi_t {\rm{ = }}\frac{{{\gamma _{t{h_t}}}}}{{\rho \left( {{b_t} - {b_l}{\gamma _{t{h_t}}} - {\varpi _2}{\gamma _{t{h_t}}}} \right)}}$ with ${b_t} > \left( {{b_l} + {\varpi _2}} \right){\gamma _{t{h_t}}}$.

Finally, combining \eqref{the expression of Theta1}, \eqref{the expression of Theta2} and \eqref{the expression of Theta3}, we can obtain \eqref{OP derived for x2}. The proof is complete.

\bibliographystyle{IEEEtran}
\bibliography{mybib}

\begin{thebibliography}{10}
\providecommand{\url}[1]{#1}
\csname url@samestyle\endcsname
\providecommand{\newblock}{\relax}
\providecommand{\bibinfo}[2]{#2}
\providecommand{\BIBentrySTDinterwordspacing}{\spaceskip=0pt\relax}
\providecommand{\BIBentryALTinterwordstretchfactor}{4}
\providecommand{\BIBentryALTinterwordspacing}{\spaceskip=\fontdimen2\font plus
\BIBentryALTinterwordstretchfactor\fontdimen3\font minus
  \fontdimen4\font\relax}
\providecommand{\BIBforeignlanguage}[2]{{%
\expandafter\ifx\csname l@#1\endcsname\relax
\typeout{** WARNING: IEEEtran.bst: No hyphenation pattern has been}%
\typeout{** loaded for the language `#1'. Using the pattern for}%
\typeout{** the default language instead.}%
\else
\language=\csname l@#1\endcsname
\fi
#2}}
\providecommand{\BIBdecl}{\relax}
\BIBdecl

\bibitem{Yuanwei2017Proceeding}
Y.~Liu, Z.~Qin, M.~Elkashlan, Z.~Ding, A.~Nallanathan, and L.~Hanzo,
  ``Nonorthogonal multiple access for {5G} and beyond,'' \emph{Proceedings of
  the IEEE}, vol. 105, no.~12, pp. 2347--2381, Dec. 2017.

\bibitem{Ding2017Mag}
Z.~Ding, Y.~Liu, J.~Choi, Q.~Sun, M.~Elkashlan, C.~L. I, and H.~V. Poor,
  ``Application of non-orthogonal multiple access in {LTE} and {5G} networks,''
  \emph{{IEEE} Commun. Mag.}, vol.~55, no.~2, pp. 185--191, Feb. 2017.

\bibitem{Islam7676258}
S.~M.~R. Islam, N.~Avazov, O.~A. Dobre, and K.~s.~Kwak, ``Power-domain
  non-orthogonal multiple access (noma) in 5{G} systems: Potentials and
  challenges,'' \emph{{IEEE} Commun. Surveys. Tutorials}, vol.~19, no.~2, pp.
  721--742, Sec. 2017.

\bibitem{QinNOMA}
Y.~Cai, Z.~Qin, F.~Cui, G.~Y. Li, and J.~A. McCann, ``Modulation and multiple
  access for 5{G} networks,'' \emph{{IEEE} Commun. Surveys. Tutorials},
  vol.~PP, no.~99, pp. 1--1, 2017.

\bibitem{Cover1991Elements}
T.~M. Cover and J.~A. Thomas, \emph{Elements of information theory}, 6th~ed.,
  Wiley and Sons, New York, 1991.

\bibitem{Ding2014Cooperative}
Z.~Ding, M.~Peng, and H.~V. Poor, ``Cooperative non-orthogonal multiple access
  in 5{G} systems,'' \emph{{IEEE} Commun. Lett.}, vol.~19, no.~8, pp.
  1462--1465, Aug. 2015.

\bibitem{Liu7445146SWIPT}
Y.~Liu, Z.~Ding, M.~Elkashlan, and H.~V. Poor, ``Cooperative non-orthogonal
  multiple access with simultaneous wireless information and power transfer,''
  \emph{{IEEE} J. Sel. Areas Commun.}, vol.~34, no.~4, pp. 938--953, Apr. 2016.

\bibitem{Men7454773}
J.~Men, J.~Ge, and C.~Zhang, ``Performance analysis of non-orthogonal multiple
  access for relaying networks over {N}akagami-$m$ fading channels,''
  \emph{{IEEE} Trans. Veh. Technol.}, to appear in 2016.

\bibitem{Yue2016Non}
X.~Yue, Y.~Liu, S.~Kang, A.~Nallanathan, and Z.~Ding, ``Outage performance of
  full/half-duplex user relaying in {NOMA} systems,'' in \emph{IEEE Proc. of
  International Commun. Conf. (ICC)}, Paris, FRA, May. 2017, pp. 1--6.

\bibitem{Shannon1961Two}
C.~E. Shannon, ``Two-way communication channels,'' \emph{in Proc. 4th Berkeley
  Symp. Math. Stat and Prob.}, vol.~1, pp. 611--644, 1961.

\bibitem{Hyadi7004056}
A.~Hyadi, M.~Benjillali, and M.~S. Alouini, ``Outage performance of
  decode-and-forward in two-way relaying with outdated {CSI},'' \emph{{IEEE}
  Trans. Veh. Technol.}, vol.~64, no.~12, pp. 5940--5947, Dec. 2015.

\bibitem{Li7778253}
C.~Li, B.~Xia, S.~Shao, Z.~Chen, and Y.~Tang, ``Multi-user scheduling of the
  full-duplex enabled two-way relay systems,'' \emph{{IEEE} Trans. Wireless
  Commun.}, vol.~16, no.~2, pp. 1094--1106, Feb. 2017.

\bibitem{Ding6868214}
Z.~Ding, Z.~Yang, P.~Fan, and H.~V. Poor, ``On the performance of
  non-orthogonal multiple access in 5{G} systems with randomly deployed
  users,'' \emph{{IEEE} Signal Process. Lett.}, vol.~21, no.~12, pp.
  1501--1505, Dec. 2014.

\bibitem{Yuanwei2017JSAC}
Y.~Liu, Z.~Qin, M.~Elkashlan, A.~Nallanathan, and J.~A. McCann,
  ``Non-orthogonal multiple access in large-scale heterogeneous networks,''
  \emph{{IEEE} J. Sel. Areas Commun.}, vol.~35, no.~12, pp. 2667--2680, Dec.
  2017.

\bibitem{RelaySharing7819537}
M.~F. Kader, M.~B. Shahab, and S.~Y. Shin, ``Exploiting non-orthogonal multiple
  access in cooperative relay sharing,'' \emph{{IEEE} Commun. Lett.}, to appear
  in 2017.

\bibitem{Liu2016TVT}
Y.~Liu, Z.~Ding, M.~Elkashlan, and J.~Yuan, ``Non-orthogonal multiple access in
  large-scale underlay cognitive radio networks,'' \emph{{IEEE} Trans. Veh.
  Technol.}, vol.~65, no.~12, pp. 10\,152--10\,157, Dec. 2016.

\bibitem{Yuanwei2017TWC}
Y.~Liu, Z.~Qin, M.~Elkashlan, Y.~Gao, and L.~Hanzo, ``Enhancing the physical
  layer security of non-orthogonal multiple access in large-scale networks,''
  \emph{{IEEE} Trans. Wireless Commun.}, vol.~16, no.~3, pp. 1656--1672, Mar.
  2017.

\bibitem{Nadarajah6831670}
S.~Nadarajah, ``A review of results on sums of random variables,'' \emph{Acta
  Appl. Math.}, vol. 103, no.~2, pp. 131--141, Sep. 2008.

\end{thebibliography}

\end{document}